\documentstyle[epsf]{l-aa}
\begin{document}
\thesaurus{6(8.15.1;8.22.1; 11.13.1)}
\title{{\sl Letter to the Editor \hspace{11.5cm}} The possible effects of an 
unusual resonance in very long period Cepheids}
\author{E. Antonello}
\institute{Osservatorio Astronomico di Brera, Via E.~Bianchi 46,
       I--23807 Merate, Italy ~(elio@merate.mi.astro.it)}
\date{ Received date; accepted date }
\maketitle
\markboth{E. Antonello: Very long-period Cepheids}
{E. Antonello: Very long-period Cepheids}

\begin{abstract}
The shape of the radial velocity and light curves of 24 long-period 
($30 \leq P \leq 134$ d) Cepheids in the Magellanic Clouds shows 
a progression with the period. The sequences of the radial 
velocity and light curves are based only on a small sample of stars; 
however, evident changes of the shape can be seen in Cepheids with period 
between 90 and 134 d. The Fourier parameter--period diagrams for the 
radial velocity curves show trends which remind in part those of Cepheids with 
period near 10 d. The plausible interpretation is a resonance, 
probably $P_0/P_1=2$ between the fundamental and the first overtone mode.
The possible importance of this phenomenon for the study of 
stellar structure and evolution in relatively far galaxies is emphasized
\footnote {Tables 2 and 3 are only available in electronic form
at the CDS via anonymous ftp to cdsarc.u-strasbg.fr (130.79.128.5)}.

\end{abstract}


\section{Introduction}

\begin{table*}
\caption[]{List of the analyzed long-period Cepheids in Magellanic Clouds}
\begin{flushleft}
\begin{tabular}{lllllllllll}
\hline\noalign{\smallskip}
HV & $P_L$ & $N_L$ & Ref$_L$  & ord$_L$ & $\sigma_L$ (mag) & $P_{RV}$ & 
$N_{RV}$ & Ref$_{RV}$   & ord$_{RV}$ & $\sigma_{RV}$ (km~s$^{-1}$)  \\
904  L  & 30.400  &  58 & 10      & 6 & .034 &        &    &   &   &     \\
1002 L  & 30.4694 &  60 & 1 - 6   & 7 & .032 &        &    &   &   &     \\ 
899  L  &         &     &         &   &      & 31.027 & 49 & 7 & 7 & 1.7 \\
2294 L  & 36.530  &  98 & 2 - 6   & 8 & .032 &        &    &   &   &     \\
879  L  & 36.817  &  31 & 4       & 5 & .048 & 36.782 & 43 & 7 & 6 & 1.4 \\
909  L  & 37.5828 &  61 & 1 - 4   & 4 & .047 & 37.510 & 64 & 7 & 5 & 1.1 \\
11182 S & 39.1941 &  44 & 2,3,5   & 3 & .039 &        &    &   &   &     \\
2257 L  &         &     &         &   &      & 39.294 & 48 & 7 & 6 & 1.7 \\
2195 S  & 41.7988 &  39 & 1,2,5   & 5 & .049 &        &    &   &   &     \\
2338 L  & 42.2153 &  55 & 2,4,5   & 7 & .048 & 42.184 & 59 & 7 & 8 & 1.4 \\
837  S  & 42.739  &  55 & 2 - 5   & 6 & .037 & 42.673 & 86 & 9 & 10 & 1.6 \\
877 L   & 45.107  &  78 & 2 - 5   & 3 & .041 &        &    &   &   &     \\ 
900 L   & 47.5417 &  62 & 1 - 5   & 5 & .039 & 47.544 & 55 & 8 & 6 & 1.2 \\
2369 L  & 48.3311 & 108 & 1 - 6   & 7 & .034 &        &    &   &   &     \\
824 S   & 65.8306 &  62 & 1,4     & 5 & .030 & 65.755 & 40 & 8 & 5 & 1.7 \\
11157 S &         &     &         &   &      & 69.06  & 62 & 9 & 5 & 1.2 \\
834 S   & 73.399  & 106 & 2 - 6   & 7 & .035 & 73.648 & 53 & 8 & 7 & 1.3 \\
2827 L  & 78.858  &  55 & 3,4     & 3 & .026 & 78.626 & 43 & 7 & 4 & 1.3 \\
829 S   &         &     &         &   &      & 85.577 & 45 & 8 & 6 & 1.0 \\
5497 L  & 99.078  &  67 & 3 - 6   & 3 & .026 & 99.040 & 41 & 8 & 3 & 1.2 \\
2883 L  & 109.277 &  58 & 2 - 4,6 & 5 & .032 & 109.071 & 52 & 8 & 3 & 2.1 \\
2447 L  & 117.941 &  68 & 3 - 6   & 3 & .022 & 118.841 & 39 & 8 & 2 & 1.3 \\
821 S   & 127.490 & 117 & 1 - 6   & 4 & .040 & 127.083 & 50 & 8 & 4 & 1.4 \\
883 L   & 133.893 &  98 & 1 - 6   & 4 & .045 & 134.75  & 92 & 9 & 6 & 2.7 \\ 
\noalign{\smallskip}
\hline
\end{tabular}
\end{flushleft}
Ref.: 1. Gascoigne \& Kron (1965); 2. Madore (1975); 3. Van Genderen 
(1977, 1983); 
4. Martin \& Warren (1979); 5. Eggen (1977); 6. Freedman et al. (1985); 
7. Imbert et al. (1985); 8. Imbert et al. (1989); 9. Imbert (1994);
10. Sebo \& Wood (1995).
\end{table*}
The longer period Cepheids are particularly important in the context
of the primary distance scale because they are bright enough to be
visibile at a great distance. However, due to their low number, they
have been poorly studied both observationally and theoretically.
Simon \& Kanbur (\cite{sk}) considered 50 Cepheids in Galaxy and
IC 4182 with period $P$ less than 70 d, compared them with 
hydrodynamical pulsation models and concluded that a detailed comparison
between theory and observations must await a more extensive and
accurate sample of observed stars. Antonello \& Morelli (\cite{am})
analyzed all the available photometric $V$ data of galactic Cepheids with 
period less than 70 d looking for possible resonance effects; they noted some
small features in the Fourier parameters--period diagrams which
were ascribed tentatively to expected resonances. Aikawa \& Antonello
(\cite{aa}) tried to reproduce these observations with nonlinear
models, but their conclusion was that the increasing 
nonadiabaticity of the pulsation with period probably reduces the 
effectiveness of resonance mechanisms. Finally, Simon \& Young (\cite{sy})
studied long period Cepheids in the period range $10 \leq P \leq 50$ d
in Magellanic Clouds looking for galaxy-to-galaxy differences in
the Cepheid distributions.
Resonances between pulsation modes, which were studied essentially in 
Cepheids with $P$ less than about 30 d, represent a powerful comparison tool 
between observations and theoretical model predictions, because they affect 
the shape of the curves of pulsating stars in specific period ranges.
The comparison of the Fourier parameters of observed and theoretical
light and radial velocity curves allows to probe the stellar 
interior and to put constraints on the stellar physical parameters.
After the work of Simon \& Lee (\cite{sl}) on the resonance $P_0/P_2=2$ 
at $P_0 \sim 10$ d between the fundamental and second overtone
mode in classical bump Cepheids, several papers by various authors were 
devoted to this topic, from both the observational and theoretical point of 
view. For example, Buchler \& Kovacs (\cite{bk}) and Moskalik \& Buchler 
(\cite{mb}) studied the general effects of 2:1 and 3:1 resonances in 
radial stellar pulsations and discussed the possible astrophysical 
implications, Petersen (\cite{pe}) discussed the possible
two- and three-mode resonances in Cepheids, and Antonello (\cite{an})
looked for the expected effects in short period Cepheids.  
Recent reviews on galactic and Magellanic Cloud Cepheids 
pulsating in fundamental and first 
overtone mode and on the problems raised by the comparison with the
pulsational models are those by Buchler (\cite{bu}) and Beaulieu \& Sasselov 
(\cite{bs}). The resonance effects in Magellanic Cloud Cepheids
cannot be reproduced by models constructed using current input physics
and reasonable mass--luminosity relations; in particular the case 
of first overtone Cepheids characterized by $P_1/P_4=2$ (Antonello 
et al. \cite{apr}) has proven to be rather difficult for theorists. 
We mention in passing also the 
recent resonance $P_2/P_6=2$, studied in the models of hypothetical second 
overtone mode Cepheids (Antonello \& Kanbur \cite{ak}). 

In the present work we have considered the long-period Cepheids in
Magellanic Clouds, with available photometric and radial velocity data
which were suitable for Fourier decomposition.
The initial purpose of the work was simply to extend 
the comparison between theory and observations to Cepheids with the
longest known periods, but the probable discovery of a new resonance
effect suggested to publish the present Letter in advance of 
the comparison with the hydrodynamical models (Antonello \& Aikawa, 
in preparation).

\section{Data Analysis}

The Cepheids with available data for a reliable analysis are reported 
in Table 1, where the subscript '$L$' refers to the light curve data
and the subscript '$RV$' to the radial velocity data. The stars are
identified with the Harvard Variable number, while the letters L and S
indicate Large and Small Magellanic Cloud, respectively. The other columns
give the period, the number of data points, the sources of the data, the
order of Fourier fit and the standard deviation of the fit for both light
and radial velocity curves. 
The time interval of all the photometric ($V$ magnitude) data is quite long 
for each star, usually
about 22 years, and during this interval the period change (probably related
to evolution) is significant, therefore it was not always 
possible to use all of the available data for the Fourier decomposition. 
In particular, for HV 5497 and HV 2447 only the observations
in the time interval between JD 2442300 and 2442900 were used.
The best photometric period was derived for each star, while
for the radial velocity data the adopted period was 
essentially the same as reported in the literature. The photometric
and radial velocity periods are usually different, because of the different
observing dates. See e.g. van Genderen (1983) for a discussion of period
changes in Magellanic Cloud Cepheids.
The adopted formula for the Fourier decomposition was
\begin{equation}
V=V_0 + \sum A_i ~ \cos [2{\pi}if(t-T_0) + \phi_i],
\end{equation}
{\it for both light and radial velocity curves}; in some data sets few
deviating points were discarded. The Tables 2 and 3 with the Fourier 
parameters, that is phase differences $\phi_{i1}=\phi_i-i\phi_1$ and 
amplitude ratios $R_{i1}=R_i/R_1$, are available at CDS.
For some stars with period between 30 and
50 d, such as HV2195, the fitted curve shows unphysical 'wiggles',
a defect which is typical of curves with steep rising branch (Antonello
\& Morelli \cite{am}). The criteria adopted for determining the best
fit were similar to those of previous works; usually these criteria yield 
different orders of best fit for different stars.  However, for stars with 
good phase coverage as in the present case, the lower order Fourier parameters 
do not change significantly when the fit is truncated at different 
higher orders (see Antonello et al. \cite{apr}).

\section {Discussion}
After examining the Fourier parameters we suspected the existence of a 
Hertzsprung--type progression for long-period Cepheids. 
In Fig. 1 we have plotted the radial velocity and light curves 
of the Cepheids with period longer than 45 d. It is possible to see
that the velocity curve differs from the 'normal' shape at
$P \sim 90$ d, it becomes progressively more symmetric and then takes again 
the 'normal' shape after $P \sim 130$ d. The light curves tend to become
more symmetric with increasing period, and between 90 and 134 d 
the shape changes near the maximum, with the possible
presence of a small bump and flat or secondary maximum.
The low order Fourier parameters are plotted in Fig. 2 and compared
with those of long-period Cepheids in the Galaxy. The data for the  galactic
Cepheids were taken from Kovacs et al. (\cite{kk}), Aikawa \& Antonello
(\cite{aa}) and Antonello \& Morelli (\cite{am}). There is a scatter
or change of phase differences $\phi_{i1}$ of the radial velocity curves
in the period range 90 - 134 d, while the $\phi_{21}$ values
of light curves are quite uniform and the $\phi_{31}$ values are scattered. 
In the same period range the amplitude 
ratios $R_{i1}$ are rather small, both for radial velocity and light
curves. These results remind in part what occurs in fundamental mode Cepheids 
with $P \sim 10$ d and in first overtone mode Cepheids with $P \sim 3$ d;
the main difference is the uniformity of $\phi_{21}$ values of the light 
curves in the present case.

Before offering the possible interpretation, some remarks are needed:
a) the number of stars in our sample is poor, and we have not discriminated
between SMC and LMC Cepheids; b) the accuracy of the 
photometric measurements is not very high and the problems related to the 
period changes cannot be avoided when selecting the data set for the 
analysis, if the observations span many years; c) the CORAVEL radial
velocity data were obtained in a short time span (less than five years),
but three Cepheids, namely HV 837, HV 11157 and HV 883, are binary, and their
pulsation curves were derived by Imbert (1994) by correcting for the
orbital motion.
In spite of these warnings, we think the progression of the curves
is real and it is related to a resonance mechanism. The linear adiabatic
models indicate $P_0/P_1=2$ between 
the fundamental and the first overtone mode as a possible candidate. 
Some years ago, Petersen (\cite{pe}) discussed this theoretical case 
using the old opacities, 
and suggested that the resonance center should be expected 
at $P_0 \sim 150$ d. As a matter of fact, the adiabatic models seem to 
indicate that the lower overtones tend to satisfy almost
simultaneously the relation $P_0 = (i+1)P_i$, or, in other words, their
frequencies tend to be coincident with the harmonics of the fundamental 
mode frequency. However, according to the linear nonadiabatic
models these resonances should not occur in the observed
period range, since the strong nonadiabaticity gives very different periods 
and period ratios from adiabatic model results
(Aikawa, private communication). 

From the comparison of galactic and Magellanic Cloud Cepheids it is 
possible to note that, even if rather scattered, the distribution of 
amplitude ratio values differs according to the galaxy: in the 
Magellanic Clouds, for $P > 30$ d, the $R_{i1}$ values can be larger than in 
the Galaxy. This result is probably related to the analogous differences of 
pulsation amplitude among Cepheids in different galaxies, studied for 
example by van Genderen (\cite{vg3}). The low number of stars do not allow to
study possible differences between LMC and SMC.

\begin{figure}
\epsfysize=13truecm
\epsffile[65 150 420 670]{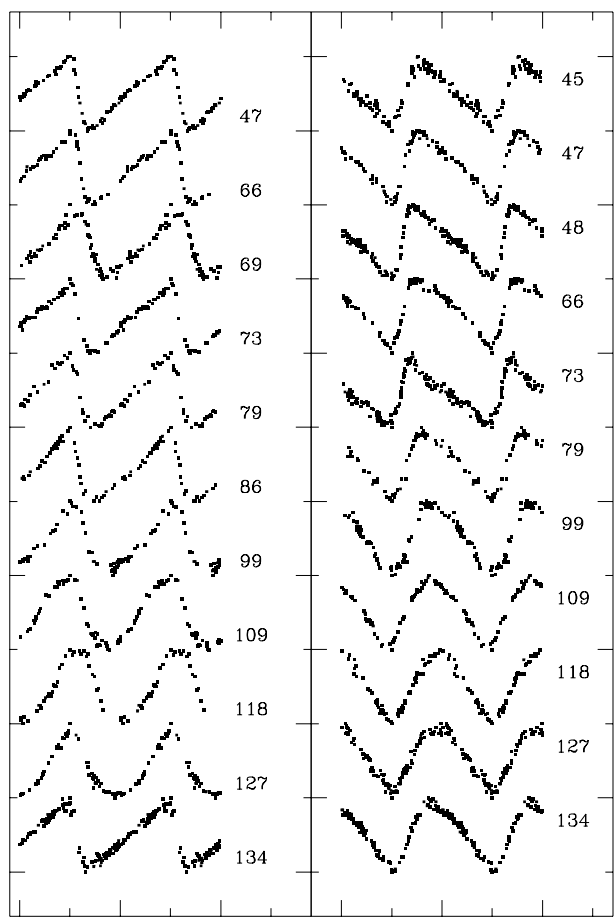}
\caption[ ]{Sequences of radial velocity ({\em left panel}) and light
({\em right panel}) curves of long-period Cepheids in Magellanic Clouds;
the periods are reported.}  
\end{figure}

\begin{figure}
\epsfysize=9.5truecm
\epsffile[50 150 380 720]{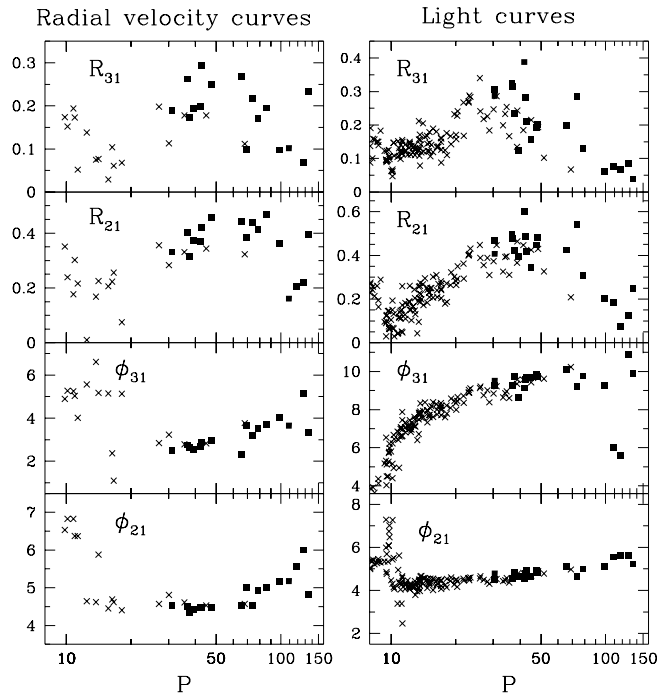}
\caption[ ]{Low order phase differences $\phi_{i1}$ and amplitude ratios
$R_{i1}$ of radial velocity ({\it left panel}) and light 
({\it right panel}) curves of long-period Cepheids in
Magellanic Clouds ({\it filled squares}) and Galaxy ({\it crosses})
}  
\end{figure}

\section{Conclusion}

The analysis of the radial velocity and light curves of the long-period
Cepheids in the Magellanic Clouds indicates the presence of a progression
which we interpret tentatively as an effect of the resonance $P_0/P_1=2$ 
between the fundamental and the first overtone mode.
Since the longest period Cepheids are also the brightest, the present 
results could be of some importance for the study of stellar structure and 
evolution in far galaxies, because the stars with $P \sim $ 100 d 
($M_V \sim -7$ mag) are about three magnitudes brighter than those at 10 d, 
in which the well known resonance $P_0/P_2=2$ is occurring. 
The disadvantage is the low number of such stars. Few Cepheids 
with $P > 80$ d have been found in relatively nearby galaxies 
(NGC 6822, IC 1613, NGC 300; see e.g. Madore \cite{mm}), while in the 
Magellanic Clouds there are just a few of stars in comparison 
with a total of some thousand Cepheids. Presently the Hubble Space 
Telescope Key Project on 
the Extragalactic Distance Scale is optimized for the detection of 
Cepheids with period between 3 and 60 d (e.g. Ferrarese et al. \cite{fe}), 
therefore it is not possible to derive reliable conclusions about the number 
of long-period stars. We just note that in NGC 925 
Silbermann et al. (\cite{sil}) found 4 stars with probable $P > 80$ d 
over a total of 80 Cepheids. 

Assuming that a sufficient number of such stars is detected and our 
interpretation is correct, the comparison of the observed resonance effect
with nonlinear model predictions will allow to test the input physics and 
put constraints on the physical parameters of the stars in relatively far
galaxies in the same way as it is occurring for the Galaxy and Magellanic
Cloud Cepheids with shorter periods.

\begin{acknowledgements}
Thanks are due to T. Aikawa and L. Mantegazza for comments and discussions.
This research has made use of the McMaster Cepheid Photometry and Radial 
Velocity Data Archive (http://www.physics.mcmaster.ca/Cepheid).
\end{acknowledgements}

\end{document}